\documentclass[british]{spie}
\usepackage[T1]{fontenc}
\usepackage[latin9]{inputenc}
\usepackage[letterpaper]{geometry}
\usepackage{float}
\usepackage{graphicx}
\usepackage{esint}

\makeatletter

\providecommand{\tabularnewline}{\\}

\makeatother

\usepackage{babel}

\begin{document}

\title{Study of tunable resonances in laser beam divergence and beam deflection}

\author{A. K\H{o}h\'azi-Kis, J. Klebniczki, M. G\"orbe and P. Nagy\skiplinehalf
Kecskemt College, Faculty of Mechanical Engineering and Automation,
H6000, Izski t. 10., Hungary}

\authorinfo{E-mail: kohazi-kis.ambrus@gamf.kefo.hu}
\maketitle
\begin{abstract}
New, fundamental resonant properties of laser resonators are theoretically
predicted and experimentally demonstrated. These resonances occur
either in the time dependence of the beam width and that of beam radius
of curvature of the wavefront or in the time dependent pointing and
position stability of the output light beam of a laser resonator.
The resonant frequency can be tuned continuously from zero to the
round-trip frequency in the first case; and from zero to the half
of the round-trip frequency in the second case, by for example, moving
one of the mirrors of the resonator. In both cases besides a resonant
frequency its complementary frequency to the round-trip frequency
is also resonant, and their shifted frequencies by multiples of the
round-trip frequency are also resonant. In our experimental demonstration
we measured the radiofrequency noise spectrum of the output laser
beam, that was partially blocked by a knife-edge. We observed increased
noise at the theoretically predicted frequencies. Similar resonances
are predicted either in the time dependent pulse-width and phase modulation
or time jitter and the central frequency of the ultrashort light pulses
of the mode-locked lasers because of the analogy between the space
description of the light beams and the time-description of the light
pulses.
\end{abstract}


\section{Introduction}

\label{sec:Introduction}

Investigating laser noise is of high interest, both in the context
of fundamental physics and for a variety of laser applications \cite{Siegman-1986,Sennaroglu-2007}.
The obtained understanding has also brought significant benefits for
the further development of lasers. The noise of lasers can also provide
us useful information about the laser alignment as in the case of
CEO-frequency measurement via heterodyning different harmonics of
the mode-locked laser spectrum \cite{Ye-2005}. The knowledge of the
fluctuations, noises in laser resonators are also important for understanding
the build up of the pulsed modes in lasers \cite{Lin-2003}.

In this paper we study radiofrequency resonances of lasers. We predict
and experimentally demonstrate two tunable resonances of continuous-wave
laser beams. The properties of these resonances are derived from the
paraxial beam propagation equations assuming real ABCD round-trip
matrices in our calculations for the sake of simplicity.

We think that the introduced resonances either have not been observed,
or have been observed but their nature was not recognized \cite{Ye-2005,Helbing-2003,Hong-2004}
or have been observed only in second order effects when the fluctuations
affect also the power of the output laser beam \cite{Lin-2003}.

The self-consistent equation for the complex radius of curvature of
light beams was perturbed in chapter 21 of Siegman's book \cite{Siegman-1986}.
The perturbed complex radius of curvature after a round trip becomes
perturbed with the same extent of deviation and the complex phase
of the deviation changes by a phase angle that depends only on the
half of the trace of the round-trip matrix. The resultant oscillations
are said to damp out due to diffraction filtering in sufficient number
of round trips\cite{Siegman-1986}. We repeat this short deduction
in Sec. \ref{sec:Beam-perturbation} just for completeness. We show,
however, in Sec. \ref{sec:Resonant-behavior} that, because of that
phase shift, laser resonators show resonant sensitivity to perturbations
from the surroundings and from the pump laser. This tunable sensitivity
cause the beam width and the radial curvature of the phase front of
the light beam of a laser to fluctuate with a frequency that depends
only on the trace of the overall round-trip ABCD matrix of the laser
resonator.

We show in Sec. \ref{sec:Beam-deflection} that if the direction of
propagation of the self-consistent light beam of a laser resonator
is perturbed then there is an also complex parameter which describes,
depending on its complex phase, propagation direction and/or transversal
position deviation. It turns out in chapter \ref{sec:Beam-deflection}
that the complex phase of this complex parameter is changed by a phase
angle that depends also on trace of the overall round-trip ABCD matrix
of the laser resonator. In chapter \ref{sec:Resonant-behavior} the
general deduction is applicable also in this case. There is a tunable
resonance of the fluctuation of the deviation angle and transversal
position of the laser beam.

In Sec. \ref{sec:Resonant-behavior} we also point out that there
are two similar resonances in the noise of the time domain parameters
of ultrashort laser pulses because of the analogy between the spatial
description of the propagation of light beams and the temporal description
of the propagation of light pulses \cite{Diels-2006}.

In Sec. \ref{sec:Experiment} we report on our experimental demonstration
of the resonances of light beams. We built a longitudinally pumped
astigmatically compensated Ti:sapphire laser resonator. The output
laser beam was partially blocked by a knife edge. We found tunable
resonances in the noise of the power of the surviving laser beam.
The observed properties of the resonances correspond to the theoretically
predicted attributes.

\section{Beam-width perturbation in laser resonators}

\label{sec:Beam-perturbation}

In this section we will consider a light beam in a laser resonator
using a paraxial approximation where the complex radius of curvature
of the beam is perturbed from its self-consistent value\cite{Siegman-1986}.
The topic of this section is practically known from Ref.~1. Laser
resonators may be astigmatic, planar resonators, but for the sake
of simplicity, we deal here only with one complex radius of curvature
describing the behaviour of the light beam in either the meridional
or in the saggital direction. We deal here only with real ABCD matrices
because our main conclusions may be drawn from this simplified picture
also. We show that if the perturbation of the complex radius of curvature
is small, only the complex phase of the perturbation changes after
a round trip in the first order of perturbation calculation. The value
of the phase change depends only on the trace of the round-trip ABCD
matrix.

The self-consistent complex radius of curvature of the laser beam
in a resonator is determined from the following equation:

\begin{equation}
q=\frac{A\, q+B}{C\, q+D}\qquad,\label{eq:SelfCons}\end{equation}
where $q$ is a complex radius of curvature of the light-beam at a
position inside the laser resonator; $A$, $B$, $C$ and $D$ are
the round-trip ABCD matrix elements \cite{Siegman-1986}. The solution
of the equation (\ref{eq:SelfCons}) can be written as

\begin{equation}
q=\frac{A-D}{2\, C}\pm\mathrm{j\:}\frac{\sqrt{1-\left(\frac{A+D}{2}\right)^{2}}}{C}\qquad,\label{eq:solution}\end{equation}
where $\mathrm{j}$ is the imaginary unit ($\mathrm{j}^{2}=-1$);
the alternative sign should be chosen to ensure the imaginary part
of $q$ to be positive. It can be achieved only if 

\begin{equation}
\left|m\right|<1\:,\qquad m=\frac{A+D}{2}\qquad,\label{eq:condition}\end{equation}
Hereafter in this paper $m$ will be called as a stability parameter
of the resonator.

We suppose that a slightly perturbed light beam starts propagating
around the resonator. The deviation of the complex radius of curvature
of the light beam starting from the self-consistent value is denoted
by $y_{1}$. The deviation of the complex radius of curvature of the
light beam after a round trip is $y_{2}$:

\begin{equation}
q+y_{2}=\frac{A\,\left(q+y_{1}\right)+B}{C\,\left(q+y_{1}\right)+D}\qquad.\label{eq:perturb1}\end{equation}

In the first order of perturbation calculation we obtain

\begin{equation}
y_{2}=\frac{A-C\, q}{C\, q+D}\: y_{1}\qquad.\label{eq:perturb2}\end{equation}

\begin{figure}
\begin{centering}
\includegraphics[height=5.5cm]{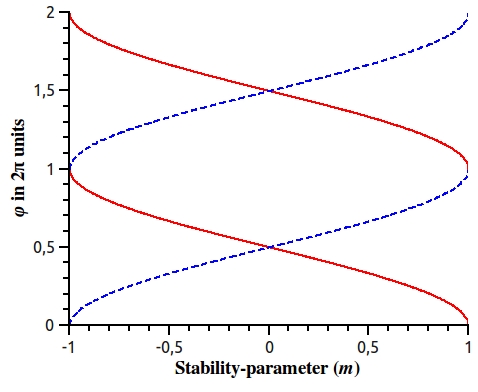}
\par\end{centering}

\caption{The phase factor as a function of the stability parameter of the laser
resonator. The continuous (red) curves are for $C<0$ and the dashed
(blue) curves are for $C>0$ (see equations (\ref{eq:solution}) and
(\ref{eq:rotation}))\label{Flo:PhaseFactor}}

\end{figure}

After using equation (\ref{eq:solution}) and the $m$ parameter defined
in (\ref{eq:condition}) and we get:

\begin{equation}
y_{2}=\left(m\mp\mathrm{j}\,\sqrt{1-m^{2}}\right)^{2}\cdot y=e^{\mathrm{j}\:\varphi\left(m\right)}\cdot y_{1}\qquad,\label{eq:rotation}\end{equation}
where $\varphi\left(m\right)$ is a complex phase-angle rotation of
one round trip (see Fig. \ref{Flo:PhaseFactor}).

It may be interesting to note that the value of the phase shift (see
equation (\ref{eq:rotation})) is equal to the double of the generalised
Gouy-phase shift\cite{Siegman-1986} of cylindrically symmetric light
beams. 

We will show in Sec. \ref{sec:Resonant-behavior} that the phase shift
determined by the equation (\ref{eq:rotation}) implies a tunable
resonant behaviour of the laser resonator.

\section{Beam-deflection in laser resonators}

\label{sec:Beam-deflection}

In this section we will consider a light beam with small angular deflection
in a laser resonator using paraxial approximation and first order
of the perturbation calculation. The propagation of a light beam through
some optical system is described by the generalised Fresnel-integral
involving the elements of the overall ABCD matrix of the given optical
system \cite{Siegman-1986}. Similarly to the treatment in Sec. \ref{sec:Beam-perturbation}
we deal only with real ABCD matrices of one dimensional beam description.

We start from the complex scalar wave function\cite{Siegman-1986}
of the self-consistent laser beam at a position inside the laser resonator
given on a transversal plane:

\begin{equation}
u\left(x_{1}\right)=E_{1}\,\exp\left(-\frac{\mathrm{j}\,\pi\, x_{1}^{2}}{\lambda\, q}\right)\qquad,\label{eq:selfcons_solution}\end{equation}
where $x_{1}$ is one of the principal transversal directions of the
paraxial laser beam, $E_{1}$ gives the amplitude and phase of the
beam, $\mathrm{j}$ is the imaginary unit, $\lambda$ is the wavelength
of the light in vacuum, and $q$ is the self-consistent complex radius
of curvature of the light beam.

An angularly deviated light beam can be described on the same transversal
plane as in equation (\ref{eq:selfcons_solution}) as follows \cite{Martinez-1986}:

\begin{equation}
u\left(x_{1}\right)=E_{1}\,\exp\left(-\frac{\mathrm{j}\,\pi\, x_{1}^{2}}{\lambda\, q}+\mathrm{j}\,2\,\beta_{1}\, x_{1}\right)\qquad,\label{eq:before_round}\end{equation}
where $\beta_{1}$ describes the angular deviation and it is supposed
to have a small value. 

After a round trip we can obtained the resultant light beam by the
generalised Fresnel-diffraction integral

\[
u\left(x_{2}\right)=e^{-\mathrm{j}\, k\, L}\,\int_{-\infty}^{+\infty}K\left(x_{2},\, x_{1}\right)\, u\left(x_{1}\right)\, dx_{1}\qquad,\]
 where $L$ is the effective path length of the laser resonator and
the kernel of the integral is \cite{Siegman-1986}

\[
K\left(x_{2},\, x_{1}\right)=\sqrt{\frac{\mathrm{j}}{B\,\lambda}}\,\exp\left[-\frac{\mathrm{j}\,\pi}{B\,\lambda}\left(A\, x_{1}^{2}-2\, x_{1}\, x_{2}+D\, x_{2}^{2}\right)\right]\qquad,\]
where $A$, $B$ and $D$ are the appropriate elements of an ABCD
matrix of the round-trip propagation. Using the Siegman-lemma \cite{Siegman-1986}:

\[
\int_{-\infty}^{+\infty}\exp\left(-a\, x^{2}-2\, b\, x\right)\, dx=\sqrt{\frac{\pi}{a}}\,\exp\left(\frac{b^{2}}{a}\right)\qquad,\]
 in the first order of the perturbation calculation we get

\begin{equation}
u\left(x_{2}\right)=E_{2}\,\exp\left(-\frac{\mathrm{j}\,\pi\, x_{2}^{2}}{\lambda\, q_{2}}+\mathrm{j}\,2\,\beta_{2}\, x_{2}\right)\qquad,\label{eq:after_round}\end{equation}
with the following three new notations

\begin{equation}
q_{2}=\frac{A\, q+B}{C\, q+D}\qquad,\label{eq:SelfCons-1}\end{equation}

\begin{equation}
\beta_{2}=\beta_{1}\,\frac{-q}{A\, q+B}\qquad,\label{eq:betak}\end{equation}

\begin{equation}
E_{2}=E_{1}\, e^{-j\, k\, L}\,\sqrt{\frac{q}{A\, q+B}}\qquad.\label{eq:GouyShift}\end{equation}

If the complex radius of curvature of the original beam given in equation
(\ref{eq:before_round}) corresponds self-consistent beam of the resonator
in the starting position and if the ABCD matrix is the round-trip
matrix of the resonator then $q_{2}=q$ (see Eq. (\ref{eq:SelfCons-1}))
just like in equation (\ref{eq:solution}).

\begin{figure}
\begin{centering}
\includegraphics[height=5.5cm]{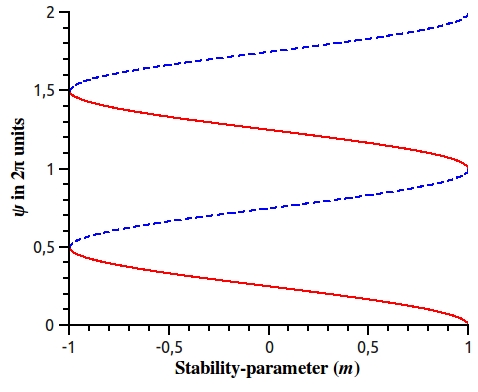}\caption{The phase factor as a function of the stability parameter of the laser
resonator. The continuous (red) curves are for $C<0$ and the dashed
(blue) curves are for $C>0$ (see equations (\ref{eq:solution}) and
(\ref{eq:fazisDev}))\label{fig:phaseDev}}

\par\end{centering}

\end{figure}

Using equations (\ref{eq:SelfCons-1}) and (\ref{eq:solution}) we
can obtain from Eq. (\ref{eq:betak})

\begin{equation}
\beta_{2}=\left(m\mp\mathrm{j}\,\sqrt{1-m^{2}}\right)\cdot\beta_{1}=e^{\mathrm{j}\,\psi\left(m\right)}\cdot\beta_{1}\quad,\label{eq:fazisDev}\end{equation}
where $\psi\left(m\right)$ is a complex phase-angle rotation of one
round trip (see Fig. \ref{fig:phaseDev}).

In equation (\ref{eq:before_round}) $\beta_{1}$ was treated as a
real variable (just to describe angular deviation), but $\beta_{2}$
in equation (\ref{eq:after_round}) turned out to be a complex quantity
generally (see equation (\ref{eq:fazisDev})). For the next round
trip $\beta_{2}$ is the starting value describing some complexized
angular deviation. We think that $\beta_{1}$ and $\beta_{2}$ should
be considered generally as a complex parameter. Depending on its complex
phase angle it can describe angular deviation and/or transversal displacement. 

It is interesting to note that the value of the phase shift (see equation
(\ref{eq:fazisDev})) is equal to the negative value of the generalised
Gouy-phase shift \cite{Siegman-1986} for cylindrically symmetric
light beams. 

From equation (\ref{eq:fazisDev}) we can conclude that after a round
trip the complex angular displacement parameter of the light beam
maintain its amplitude and changes its complex phase angle by an amount
depending on the stability parameter ($m$) of the resonator (see
equation (\ref{eq:condition})) in the first order perturbation calculation.
That is after a round trip the real and imaginary parts of the complex
angular deviation are coupled: the deviation in beam propagation direction
can be transformed to deviation in the transversal position of the
light beam, and vice versa.

We will show in Sec. \ref{sec:Resonant-behavior}. that this phase
shift determined by the equation (\ref{eq:fazisDev}) implies also
a tunable resonant behaviour of the laser resonator similarly to the
phase shift determined in Sec. \ref{sec:Beam-perturbation}.

\section{Resonant behaviour}

\label{sec:Resonant-behavior}

In this section we show that a light beam in a laser resonator (or
in any optical system with feedback) has a resonant noise sensitivity
if a complex beam parameter gets a complex phase shift per round trip.

Let us consider a complex-valued beam parameter $y$ excited by some
noise $n$. The value of $y$ at a time $t$ is the result of an evolution
from the previous round-trip as follows: 1. the parameter suffers
attenuation as multiplication by (1-$\alpha$); 2.a the parameter
propagates through the feedback loop which formally appears as multiplicatio
by a complex phase factor $e^{j\,\varphi}$; 2.b as part of the propagation,
the parameter is delayed by time interval $T$; 3. the noise added
at a time $t$ (see Fig. \ref{Flo:feedback}):

\begin{figure}
\centering{}\includegraphics[height=5cm]{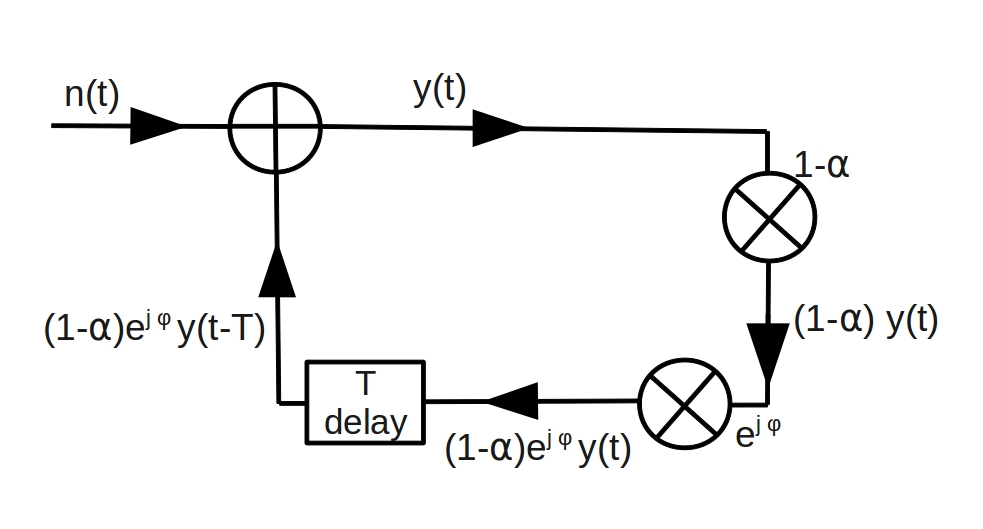}\caption{A schematic diagram of the feedback loop\label{Flo:feedback}}

\end{figure}

\begin{equation}
n\left(t\right)+\left(1-\alpha\right)\, e^{\mathrm{j}\,\varphi}\, y\left(t-T\right)=y\left(t\right)\quad,\label{eq:feedback}\end{equation}
where $\varphi$ is a constant phase value, $\left(1-\alpha\right)$
is an attenuation factor ($0<\alpha<1$) and $T$ is a constant time
delay.

After taking the Fourier-transform of equation (\ref{eq:feedback})
we get

\begin{equation}
N\left(f\right)+\left(1-\alpha\right)\, e^{j\,\varphi}\: Y\left(f\right)\, e^{j\,2\,\pi\, f\, T}=Y\left(f\right)\quad,\label{eq:Feedback}\end{equation}
where $f$ is a frequency value, $N\left(f\right)$ and $Y\left(f\right)$
are the Fourier-transform of the noise $n\left(t\right)$ and the
beam parameter $y\left(t\right)$ time functions, respectively.

\begin{equation}
\left|Y\left(f\right)\right|^{2}=\frac{\left|N\left(f\right)\right|^{2}}{1+\left(1-\alpha\right)^{2}-2\,\left(1-\alpha\right)\,\cos\left(\varphi-2\,\pi\, f/f_{0}\right)}\quad,\label{eq:ResSol}\end{equation}
where $f_{0}=1/T$ .

\begin{figure}
\centering{}\begin{tabular}{cc}
\includegraphics[height=5.5cm]{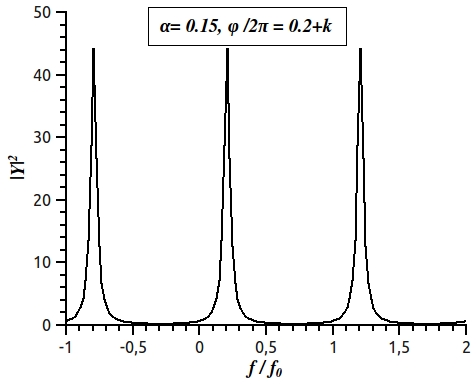} & \includegraphics[height=5.5cm]{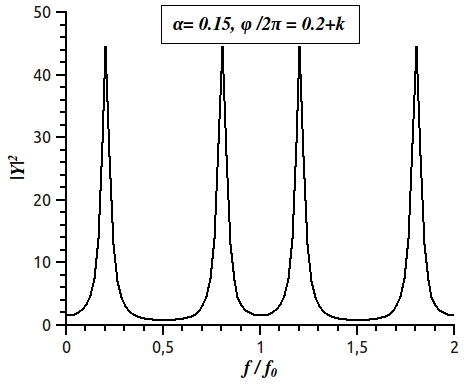}\tabularnewline
a) & b)\tabularnewline
\end{tabular}\caption{The amplitude-square spectrum of the investigated beam parameter (a)
and a real valued i.e. measurable signal (b)\label{Flo:ResSpectrum}}

\end{figure}

The amplitude-square spectrum $\left|Y\left(f\right)\right|^{2}$
shows a resonant behaviour. If we suppose the spectrum of noise to
be constant (white noise) then we get simple resonance curves (see
Fig. \ref{Flo:ResSpectrum}.a).

The resonance in the amplitude of $\left|Y\left(f\right)\right|$
occurs when the denominator is minimal in equation (\ref{eq:ResSol}),
that is 

\begin{equation}
f_{res}=f_{0}\,\left(\frac{\varphi}{2\,\pi}+k\right)\quad,\label{eq:ResFrequency}\end{equation}
where $k$ is an arbitrary integer value.

When we measure, naturally, real quantity that depends on the complex
beam parameter $y$ we would observe resonances not only at positive
frequencies given by equation (\ref{eq:ResFrequency}) but also at
the absolute value of the negative frequencies also given by equation
(\ref{eq:ResFrequency}). If a frequency is resonant then the complementary
frequency to the round-trip frequency is also resonant. Finally we
obtain noise spectra similar to the curve given by Fig. \ref{Flo:ResSpectrum}.b
instead of the curve given by Fig. \ref{Flo:ResSpectrum}.a as it
was checked with simple simulations.

\begin{figure}[H]
\begin{centering}
\begin{tabular}{cc}
\includegraphics[height=5.5cm]{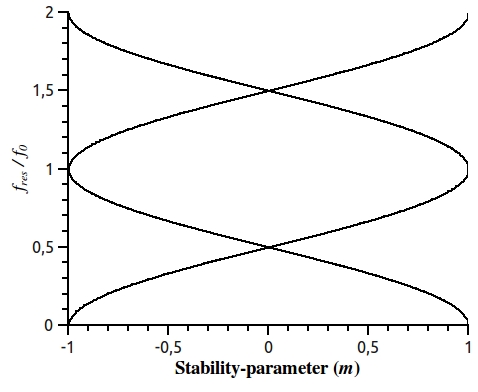} & \includegraphics[height=5.5cm]{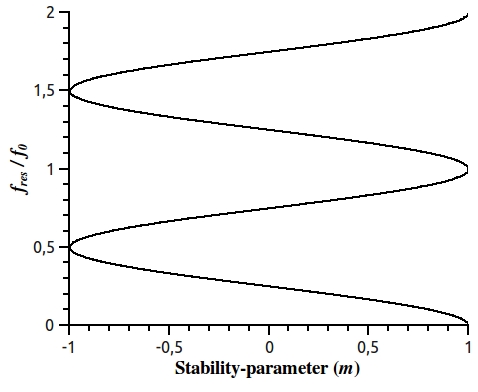}\tabularnewline
a) & b)\tabularnewline
\end{tabular}
\par\end{centering}

\caption{Resonance frequency as functions of the stability parameter: a) resonance
in the beam width fluctuation b) resonance in the angular deviation
fluctuation of the beam. The curves do continue to higher frequencies
according to equation (\ref{eq:ResFrequency}) \label{fig:Resonance-frequency}}

\end{figure}

According to our analysis of this section we can plot the resonant
behaviour in the beam width and angular deviation in Fig. \ref{fig:Resonance-frequency}.

There is an analogy between the spatial description of light beams
and the temporal description of ultrashort light pulses: their evolution
is described by appropriate two-by-two matrices \cite{Diels-2006}.
In a time domain, dispersion and self-phase modulation correspond
to spatial distance and focusing in the spatial description, respectively.
The calculation presented in Sec. re-derived in time domain \ref{sec:Beam-perturbation}
would result a tunable resonance of the duration and the phase modulation
of the laser pulses, while the calculation presented in Sec. \ref{sec:Beam-deflection}
re-derived in tima domain would result a tunable resonance of the
timing jitter and the central frequency of the laser pulses. In these
cases the tuning can be achieved by varying the dispersion and/or
the self-phase modulation in the laser resonator. We expect these
resonances to be detectable for example, in the noise of the second
harmonic signal of the output laser beam. We think that these resonances
have been detected but has been misinterpreted \cite{Hong-2004,Helbing-2003,Ye-2005}.
The authors of the cited works labelled some resonances as {}``spurious''
and interpreted them as being generated by nonlinear electronic mixing
processes. We suppose those resonances to be the type here being investigated
because of their occurrence in pairs just as in Fig. \ref{Flo:ResSpectrum}.b.

The CEO (carrier-envelope offset) phase of an ultrashort light pulse
in a laser resonator is also a complex parameter that changes a given
amount per round trip. That is why the CEO-frequency also has the
property described by equation (\ref{eq:ResFrequency}).

\section{Experiment}

\label{sec:Experiment}

\begin{figure}
\centering{}\includegraphics[height=7cm]{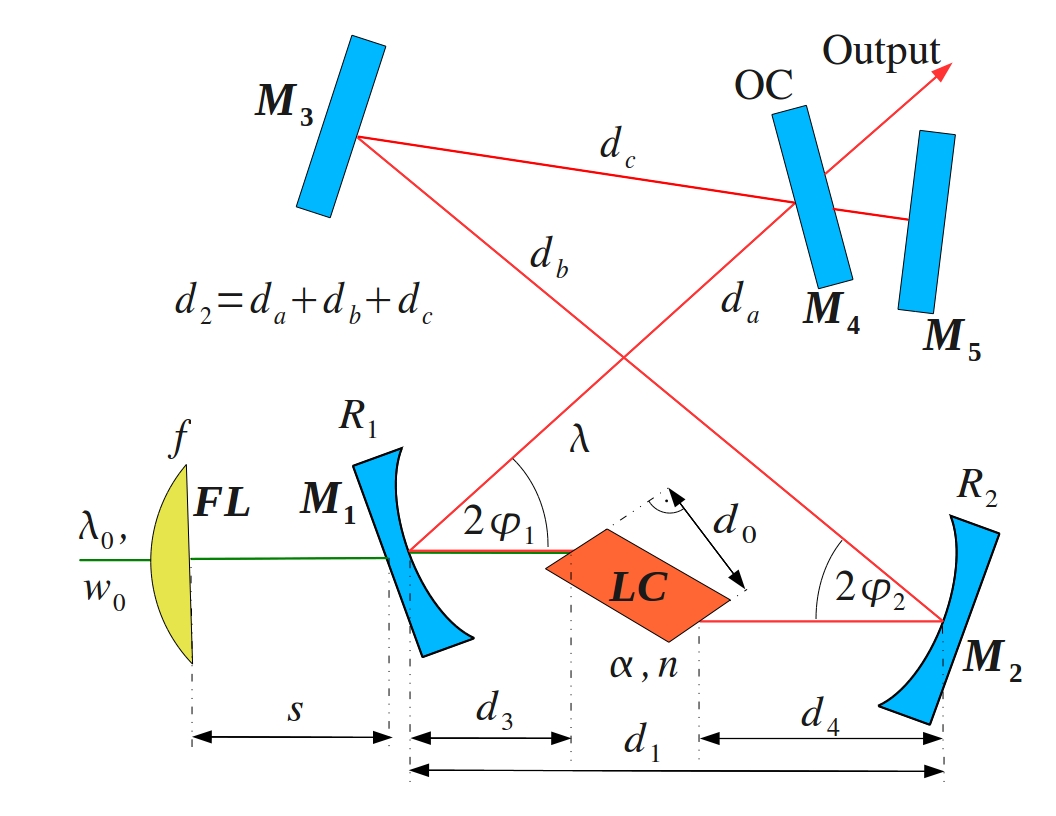}\caption{Schematic diagram of the laser resonator\label{Flo:Resonator}}

\end{figure}

We have built a longitudinally pumped astigmatically compensated Ti:sapphire
ring laser (see fig. \ref{Flo:Resonator} and table \ref{Flo:LasParam})
to demonstrate the predicted resonances of light beams. We chose ring-resonator
arrangement to realize the model described is Secs. \ref{sec:Beam-perturbation}
and \ref{sec:Beam-deflection}: the light beam in the resonator was
perturbed in the laser crystal only once a round-trip. The parameters
of our laser resonator can be seen in table \ref{Flo:LasParam}. Moreover
this laser configuration has the beneficial property that the stability
parameter ($m$) introduced in Sec. \ref{sec:Beam-perturbation} is
linearly scaled with the $d_{4}$ parameter of the laser resonator
(Fig. \ref{Flo:Resonator}). This property was used to scan the stability
range of the laser.

\begin{table}[h]
\centering{}\begin{tabular}{|c|c|c|}
\hline 
$\lambda_{0}=0.532\,\mathrm{\mu m}$ & $w_{0}=1.15\,\mathrm{mm}$ & $f=84\,\mathrm{mm}$\tabularnewline
\hline 
$\varphi_{1}=5.0^{\circ}$ & $\varphi_{2}=10.9^{\circ}$ & $R_{2}=R_{1}=75\,\mathrm{mm}$\tabularnewline
\hline 
$d_{0}=3.46\,\mathrm{mm}$ & $\alpha=0.45\,\mathrm{mm}^{-1}$ & $n=1.752$\tabularnewline
\hline 
$\lambda=800\,\mathrm{nm}$ & $d_{2}=1375\,\mathrm{mm}$ & $Refl_{OC}=92\%$\tabularnewline
\hline
\end{tabular}\caption{The parameters of the astigmatically compensated laser-resonator whose
schematic layout can be seen on Fig. \ref{Flo:Resonator}.\label{Flo:LasParam}}

\end{table}

The resonator is not a special resonator, it is optimised just for
maximal output power at 4 W pump power. The mirrors in the resonator
are simple quarter-wave stack mirrors. The laser was pumped by a Verdi
G5 laser (Coherent Inc.). The parameters of the laser resonator can
be seen in table \ref{Flo:LasParam}.

The focusing lens (FL), the laser-crystal (LC) and the $M_{2}$ mirror
were mounted on translator stages with micrometer screws. The position
of these elements originally were optimised for minimal pump power,
but during the measurements the position of the focusing lens and
the laser crystal were left unmoved. The direction of the movement
of the $M_{2}$ mirror was not perfectly parallel to the direction
of the light beams, that is why the $M_{3}$, $M_{4}$ and $M_{5}$
mirrors were adjusted for minimal pump power each time the $M_{2}$
mirror was moved. At pump laser power of 4 W we could achieve laser
action at the micrometer position of $M_{2}$ mirror between 4.00
mm and 7.80~mm values.

To detect the resonances described in Secs. \ref{sec:Beam-perturbation}-\ref{sec:Resonant-behavior}.
we set a knife-edge on a micrometer positioner with a vertical edge
into the beam path of the output laser beam at a distance of 95 cm
from the output coupling mirror ($M_{4}$). Behind the knife edge
at a distance of 15 cm a fast photodetector (Newport model 818-BB-22)
was measuring the power of the residual light beam. The residual light
beam was regularly checked to fall on the sensitive surface of the
photodetector. The signal of the photodetector was measured by an
Agilent DSO 9254 oscilloscope. We used external 50 $\Omega$ terminator
resistor. The noise was measured at a sensitivity of 1 mV/scaling
at a 2 GSample/s sampling rate measuring 32768 points a scan using
AC coupling. The data of the scans was Fourier-transformed and later
averaged by the plug-ins of the oscilloscope. For adjustment we used
64 time averaging, for measuring 1024 time averaging. The background
noise of the detecting system was between -95 dBm and -97 dBm values
in the investigated 30-250 MHz frequency range besides some background
resonances of the surroundings. (The values given in dBm units means
the ratio of the given electric signal's power on 50 $\Omega$ resistor
relative and the 1 mW value given in dB units. We checked this law
width a signal generator.)

\begin{figure}[h]
\begin{centering}
\includegraphics[height=5.5cm]{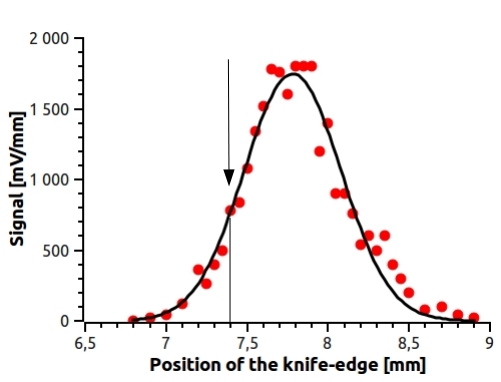}
\par\end{centering}

\caption{Shape of the output laser beam measured by a knife-edge method. \label{fig:KnifeEdge}}

\end{figure}

The shape of the laser beam was measured by the knife-edge method
(see Fig. \ref{fig:KnifeEdge}). To preclude saturation of our detection
system we used neutral grey-filter before the photodetector.

\begin{figure}[h]
\begin{centering}
\includegraphics[height=5.5cm]{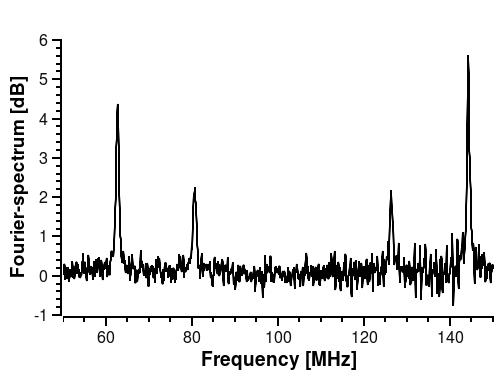}
\par\end{centering}

\caption{The spectrum of the photodetector signal at a certain mirror and knife
edge position\label{Flo:MeasResonancies}}

\end{figure}

Two pairs of resonances can be observed on Fig. \ref{Flo:MeasResonancies}
which shows a typical measured spectrum of the signal of the photodetector
above the background noise level. When the $M_{2}$ mirror was moved
the resonances were sweeping in the frequency range. The resonances
were occurring always in pairs: there was in the middle position the
half of the round-trip frequency just the way that was predicted in
Sec. \ref{sec:Resonant-behavior}.

\begin{figure}[h]
\begin{centering}
\begin{tabular}{cc}
\includegraphics[height=5.5cm]{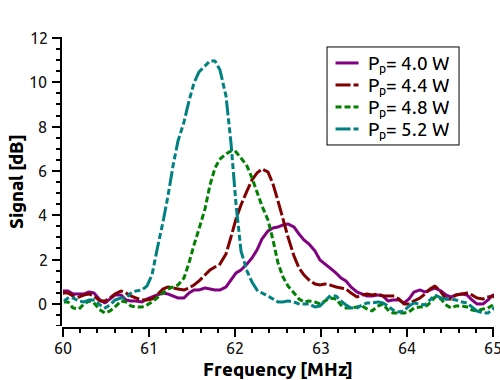} & \includegraphics[height=5.5cm]{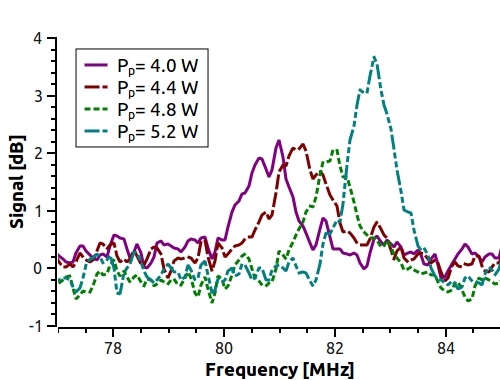}\tabularnewline
a) & b)\tabularnewline
\end{tabular}
\par\end{centering}

\caption{Pump power dependence of the resonant frequencies. Resonance of the
beam tilt angle (a) and the beam width (b) variation (see Fig. \ref{Flo:MeasResonancies})
\label{Flo:PowDependence}}

\end{figure}

Pump power dependency of the resonant frequencies are measured at
a fixed mechanical alignment (see Fig.~\ref{Flo:PowDependence}).
The change of the pump power changes the effective focal length of
the thermal lensing and the gain guiding lensing. These changes deviate
elements of the round-trip matrix from the values calculated from
a bare resonator theory. These deviations keep us from being able
to precisely determine the mirror positions from the measured resonant
frequency values. But from these deviations we may determine the strength
of self focusing and thermal lensing.

\begin{figure}[h]
\begin{centering}
\includegraphics[height=5.5cm]{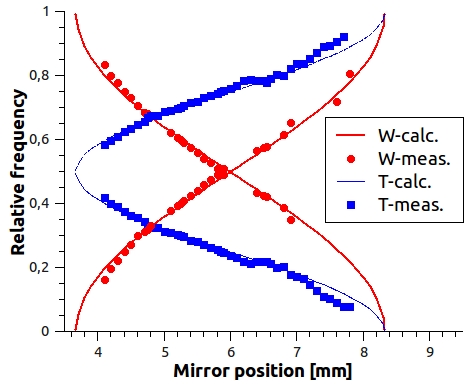}
\par\end{centering}

\caption{Measured frequencies of resonances as a function of the $M_{2}$ mirror
position\label{Flo:MeasFreqencies}}

\end{figure}

We measured the dependency of the frequencies of the resonances as
a function of the $M_{2}$ mirror position (see Fig. \ref{Flo:Resonator}).
The relative frequency means here the ratio of the measured frequency
of the resonance and the round-trip frequency of the laser resonator.
The frequencies denoted by red circles are connected to the beam-width
fluctuation (W-meas) and the frequencies denoted by blue rectangles
are connected to the beam-tilt deviation (T-meas). The lines on the
figure show fitted calculated dependencies of the resonator frequencies
(see Fig. \ref{fig:Resonance-frequency}). We measured similar resonances
in the saggital direction (with horizontal knife edge), the observed
frequencies corresponds well to the shifted effective resonator.

We checked that there is no resonance in the noise of the power of
the entire output beam at the same setup as in the case of Fig. \ref{Flo:MeasResonancies}.
To avoid saturation of our measuring system we attenuated the beam
by a reflection from a glass surface.

The resonances connected to the fluctuation of the beam width variation
was always observed at the edge of the light beam (corresponding to
high beam-stop ratio), where the DC signal was high enough (without
any attenuation of the light beam other than the effect of the knife
edge).

\begin{figure}[h]
\begin{centering}
\begin{tabular}{cc}
\includegraphics[height=5.5cm]{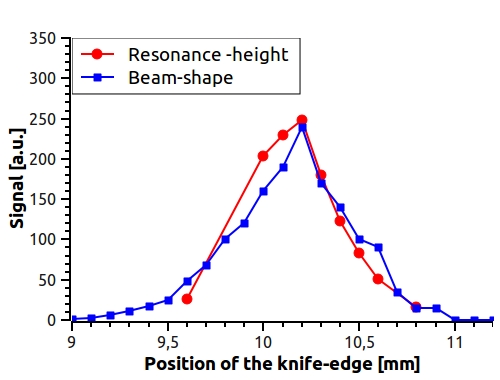} & \includegraphics[height=5.5cm]{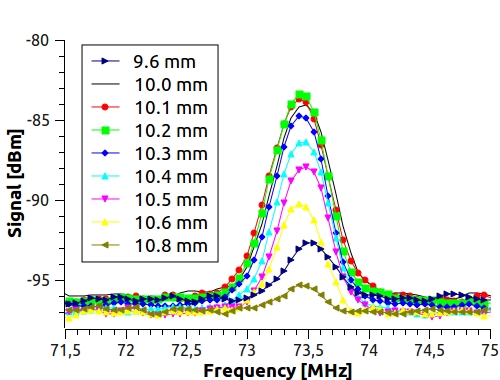}\tabularnewline
a) & b)\tabularnewline
\end{tabular}
\par\end{centering}

\caption{Measuring the beam deviation resonance with changing the position
of the knife edge (the position of the $M_{2}$ mirror was 4.60 mm).
a) The amplitude of the resonance and the beam intensity as a function
of the position of the knife edge. b) The shape of the resonances
at different positions of the knife edge \label{fig:TiltKnife}}

\end{figure}

The resonances connected to the fluctuation of the beam propagation
direction could have been observed at every position of the knife
edge provided that the measuring system did not get saturated (see
Fig. \ref{fig:TiltKnife}). To avoid saturation we applied a neutral
grey-filter before the photodetector. The amplitude of the fluctuation
is proportional to the amplitude of the light intensity at a given
position of the knife edge (see Fig. \ref{fig:TiltKnife}), as it
can be expected from the theory.

We found the same kind of resonances in the saggital direction also,
that is, when the knife with vertical edge was moved in horizontal,
transversal direction. The frequencies of the resonances was shifted
according to the shifted stability ranges in the meridional and in
the saggital directions.

\section{Summary}

We theoretically predicted and experimentally demonstrated two new,
fundamental resonances of laser resonators. These radiofrequency resonances
occur in the power noise in the output laser beam partially blocked
by a knife edge. There is a tunable resonance of the beam-width and
the radius of curvature of the output light beam, and there is an
another resonance of the angular- and the transversal positional deviation
of the light beams. The resonant frequencies can be tuned with, for
example, changing the position of one of the mirrors of the resonator.

The resonant behaviour was experimentally demonstrated with a longitudinally
pumped astigmatically compensated Ti:sapphire ring laser. Using a
knife-edge to stop most of the output beam and only its small part
was measured with a fast linear detector. The signal was observed
by a digital oscilloscope. A Fourier-transform was taken of the sample
points. The Fourier-spectrum was averaged many times to reduce its
noise. We observed resonant behaviour in the noise of the detected
signal. The observed resonances was 3-10 dB above the noise-background
of the measuring system. If we observe resonance at a frequency value
then we also observe resonance at the complement frequency up to the
round-trip frequency. The observed dependencies of the frequencies
of the resonances on the stability parameter of the laser resonator
was in good agreement with the theory. One of the possible applications
of these resonances are an approximate mapping tool of the stability
region.

Because of the analogy between the description of light beams in the
spatial domain and the description of light pulses in the time-domain
we expect two type of tunable resonances in the noise of the second
harmonic of the output laser pulse train. We expect a tunable resonance
of the pulse-duration and the phase modulation of the output light
pulse train, and we expect an another resonance of the time-jitter
and the central frequency deviation of the pulse train.

\section{Acknowledgment}

The authors wish to acknowledge the TIOP 1.3.1./07/2/2F/2009/0003
grant for financial support of this research. In addition, we would
like to give special thanks to the Rohde \& Schwarz Hungary Ltd. lending
us their oscilloscope and spectrum analyser just to check our signals
with another equipments also.\bibliographystyle{plain}
\bibliography{LasRez}

\end{document}